\documentclass[fleqn,usenatbib]{mnras}

\usepackage[english]{babel}
\usepackage[usenames]{color}

\usepackage{newtxtext,newtxmath}
\usepackage[T1]{fontenc}
\usepackage{graphicx}	
\usepackage{amsmath}	

\title[Collapse of magnetic rotating protostellar clouds]{Simulations of the isothermal collapse of magnetic rotating protostellar clouds}

\author[S. A. Khaibrakhmanov, et al.]{
Sergey Khaibrakhmanov$^{1,2}$\thanks{E-mail: khaibrakhmanov@csu.ru},
Alexander Dudorov$^{2,1}$\thanks{deceased},
Natalya Kargaltseva$^{1,2}$,
Andrey Zhilkin$^{3}$
\\
$^{1}$Ural Federal University, 51 Lenin st., Ekaterinburg 620000, Russia\\
$^{2}$Chelyabinsk State University, 129 Br. Kashirinykh st., Chelyabinsk 454001, Russia\\
$^{3}$Institute of Astronomy of the Russian Academy of Sciences, 48 Pyatnitskaya st., Moscow 119017, Russia
}

\date{Accepted 09.05.2021}
\pubyear{2021}

\begin{document}
\label{firstpage}
\pagerange{\pageref{firstpage}--\pageref{lastpage}}
\maketitle

\begin{abstract}
We investigate collapse of magnetic protostellar clouds of mass  $10$ and $1\,M_{\odot}$. The collapse is simulated numerically using the two-dimensional magnetohydrodynamic (MHD) code `Enlil'. The simulations show that protostellar clouds acquire a hierarchical structure by the end of the isothermal stage of collapse. Under the action of the electromagnetic force, the protostellar cloud takes the form of an oblate envelope with the half-thickness to radius ratio $Z/R \sim (0.20-0.95)$. A geometrically and optically thin primary disk with radius  $0.2-0.7\,R_0$ and $Z/R \sim (10^{-2}-10^{-1})$ forms inside the envelope, where $R_0$ is the initial radius of the cloud. Primary disks are the structures in magnetostatic equilibrium. They form when the initial magnetic energy of the cloud exceeds 20~\% of its gravitational energy. The mass of the primary disk is 30--80~\% of the initial mass of the cloud. The first hydrostatic core subsequently forms in the center of the primary disc. We discuss the role of primary disks in the further evolution of clouds, as well as possible observational appearance of the internal hierarchy of the collapsing cloud from the point of view of the features of the magnetic field geometry and the distribution of angular momentum at different levels of the hierarchy. 

\end{abstract}


\section{Introduction}
\label{sect:intro}

Observations in the optical, infrared and submillimeter ranges have shown that modern star formation takes place in the cold cores of molecular clouds. We will call these cores as protostellar clouds (PSCs hereafter). PSCs have typical sizes of $0.05-0.3$~pc, densities of $10^4 - 10^5$~cm$^{-3}$, masses from several tenths to tens of solar masses~\citep[e.g.][]{teixeira2005, enoch11}, temperatures of 10--20~K~\citep[e.g.][]{launhardt2013, tang2018, sadavoy2018}. Velocity gradients determined from molecular emission lines of different parts of the PSCs indicate solid-body rotation~\citep{caselli02, punanova2018}.The rotational energy is a few percent of the gravitational energy. A large-scale magnetic field with an intensity of the order of $10^{-5}$~G is found in PSCs with the help of polarization mapping and measurements of the Zeeman splitting of OH molecular lines~\citep{troland2008, li2009, crutcher2012, galametz2018}.

The gravitational collapse of the PSC leads to the formation of a protostar, which is observed as a source of infrared (IR) radiation in the center of the cloud. The earliest stage in the evolution of the protostar, when it is deeply embedded in an extended envelope of gas and dust, defined by a radiation peak in the submillimeter range, is associated with class 0 young stellar objects~\citep{andre1993}.

Observations of several class 0 young stellar objects (YSOs hereafter) in molecular emission lines and in the IR continuum have shown that young protostars are surrounded by extended flattened structures with radii from several hundred to several thousand astronomical units~\citep{ohashi97, wiseman2001, looney2007, tobin2010, lee2014}. Some of these structures have rotation profiles corresponding to spatially constant angular momentum~\citep{tobin2012, lee2019, guedel2020}. Recent observations of the same objects with the Atacama Large (sub-) Mm Array (ALMA) and Very Large Telescope (VLT) with high angular resolution showed that there are smaller compact disks inside these flattened envelopes around the protostars~\citep{jorgensen09, enoch11, tobin2020, maret2020}. Some of these compact discs exhibit rotation profiles close to the Keplerian one~\citep{murillo2013, lindberg2014, yen2017}.

It was possible to carry out polarization mapping and determine the geometry of the large-scale magnetic field in the disks for several class 0 YSOs. For example, pinched magnetic field and signs of toroidal magnetic field were found inside the disk of HH 211~\citep{lee2019}. The magnetic field in VLA 1632A is directed along the outflow axis~\citep{Hull_2014}. Molecular outflows and jets are typical features of class 0 YSOs~\citep{andre1993}. \cite{girart2006}, \cite{ Davidson_2011}, \cite{Chapman_2013} found an hourglass magnetic field in several clouds at the scales comparable to the size of the cloud.

Analysis of the above observational data shows that the dynamics of the collapsing PSCs in class 0 YSOs is characterized by the formation of compact disks embedded into large-scale flattened envelopes. 

The first numerical simulations of the collapse of PSCs have shown that the initial isothermal stage of the collapse is characterized by the formation of the flattened envelope of the collapsing cloud~\citep{nakano1979, black1982}. Based on the collapse simulations within the framework of the `inside-out' collapse model, \cite{Galli1993} stated that, during the collapse of an isothermal non-rotating singular sphere with uniform magnetic field, a pseudodisc is formed, which is a transient flattened nonequilibrium structure. In later works, any flattened structures formed at the initial stages of the collapse were called pseudodisks~\citep[e.g.][]{tomisaka2002, hennebelle2009, tsukamoto2017, Zhao18, lam2019}, although this term has no strict definition. The main problem considered in the above-mentioned papers is the so-called magnetic braking catastrophe, formulated by~\cite{galli2006}. The problem is that the frozen-in magnetic field prevents the formation of Keplerian disks during the collapse of the PSC due to the effective removal of angular momentum by torsional Alfvén waves. The solution to this problem is obviously related to the mechanisms of attenuation of the magnetic flux of the cloud, namely: diffusion of the magnetic field and/or turbulence~\citep[see recent review][]{zhao2020}. Since different teams of authors use different problem statements and numerical codes, it has not yet been possible to determine the relative role of these mechanisms. 

It is clear that the characteristics of the primary flattened structures, in particular the distribution of their angular momentum and magnetic flux, determine the conditions for the formation of Keplerian disks around protostars. However, the internal structure and characteristics of such large-scale flattened structures formed at the early stages of the collapse have not been studied in detail. It is worth noting the works of~\cite{dudorov1999mmb, dudorov2000}, who have shown that the thickness to radius ratio of the flattened structures formed in the early stages of the collapse is proportional to $\varepsilon_{\rm m}^{-1/2}$, where $\varepsilon_{\rm m}$ is the ratio of the magnetic energy of the cloud to the modulus of its gravitational energy. Based on the results of two-dimensional MHD modeling, \cite{tomisaka2002} noted that the epoch of the formation of primary flattened structures is characterized by the formation of MHD shock waves propagating into the cloud envelope and by the subsequent formation of outflows. This result confirmed the predictions of~~\cite{DudSaz1982}. Previous studies have focused on the collapse of the PSC with the mass of the Sun (see references above) or massive stars with a mass greater than  $10\,M_{\odot}$~\citep[see recent review][]{zhao2020}.

In this work, we investigate the isothermal stage of the collapse of magnetic PSCs of intermediate, $M=10\,M_{\odot}$, and solar mass in a wide range of initial thermal and magnetic energies. We focus on the formation of quasi-magnetostatic disks, which we call primary disks. The main goal of this work is to study the characteristics of the primary disks in detail since they determine the further formation and evolution of the protostar with the protostellar disk. The problem statement of the collapse of the PSC in the axisymmetric approximation, the numerical method and the parameters of the model are described in Section~\ref{sect:problem}. Section~\ref{sect:fiduc_m} analyzes the typical picture of the isothermal collapse of the magnetic PSC and determines the characteristics of the resulting hierarchical structure: the flattened envelope and the primary disk. Section~\ref{sect:evol} examines the evolution of primary disks. The main features of the formed structures, depending on the initial parameters of the cloud, are analyzed in Section~\ref{sect:param}. Section~\ref{sect:end} discusses the general picture of the isothermal collapse of the PSC, discovered in our simulations, and outlines the prospects for further research. 

\section{Problem statement}
\label{sect:problem}

Consider an uniform spherically symmetric PSC with mass of $ M_0 = 10 M_{\odot}$ or $1\,M_{\odot}$ and temperature of $T = 20$~K.
The cloud rotates rigidly and is threaded by a uniform magnetic field. The cloud is in pressure equilibrium with the external environment initially. Initial concentration $n_0$, radius $R_0$ and magnetic field strength $B_0$ of the cloud are specified with the help of dimensionless parameters $\varepsilon_{\rm t}$ and $\varepsilon_{\rm m}$, which are the ratios of the thermal and magnetic energies of the cloud to the modulus of its gravitational energy, respectively. 

We investigate the collapse of the PSC based on the system of equations of gravitational MHD in the approximation of the frozen-in magnetic field: 
\begin{eqnarray}
  \frac{\partial\rho}{\partial t} + {\bf \nabla}\cdot\left(\rho\textbf{v}\right) &=& 0, \label{Eq:Cont}\\
  \rho\left[\frac{\partial {\bf v}}{\partial t} + \left({\bf v}\cdot \nabla\right){\bf v}\right] &=& -\nabla P - \rho{\bf \nabla} \Phi \nonumber\\
  & & + \frac{1}{4\pi}\left(\nabla\times {\bf B}\right) \times {\bf B},\label{Eq:Motion}\\
  \rho\left[\frac{\partial \varepsilon}{\partial t} + \left({\bf v}\cdot {\bf \nabla}\right)\varepsilon\right] + P{\bf \nabla}\cdot{\bf v} &=& 0,\label{Eq:TotalEnergy}\\
  \frac{\partial\textbf{B}}{\partial t} &=& \nabla\times \left(\textbf{v} \times \textbf{B}\right), \label{Eq:Induction}\\
  \nabla^2\Phi & = & 4\pi G\rho,\\
  P & = & (\gamma - 1)\varepsilon\rho,
\end{eqnarray}
where $\Phi$ is the gravitational potential, $\varepsilon$ is the specific internal energy of the gas, $\gamma$ is the adiabatic index. All other values are used in standard physical notation. The ideal MHD approximation is valid for the considered initial stages of collapse, when the density does not increase by more than 4--5 orders of magnitude compared to the initial one and does not exceed $10^{12}\,\mathrm{cm}^{-3}$~\cite[see][]{nakano2002}. The gas temperature remains practically constant at this stage. We adopt the effective value of the adiabatic index~$\gamma=\gamma_{\rm{eff}} = 1.001$~to simulate the isothermal collapse, when the gas in the cloud is effectively cooled and its temperature does not change.

We use a cylindrical coordinate system  $(r, \, \varphi, \, z)$. The origin is at the center of the cloud. The $z$--axis is directed along the initial magnetic field direction ${\mathbf B}_0$. In the axisymmetric approximation, one can consider the cloud dynamics in the plane $r$ and $z$ in the region of positive $r$ and $z$. A detailed description of the model equations in cylindrical coordinates is given in a paper by~\cite{zhilkin09}. 

Numerical simulation of the collapse of the PSC is performed using the MHD code `Enlil' utilizing an adaptive moving mesh~\citep{Dud1999, dudorov1999mma, dudorov1999mmb, zhilkin09}. The system of equations of the model is solved in dimensionless form. Initial density $\rho_0$ and radius $R_0$ of the cloud are used as the density and coordinate scales, respectively. The time is measured in units of $t_0=1/\sqrt{4\pi G\rho_0}$. The corresponding scales of velocity, pressure, energy and magnetic field strength are $v_0=R_0/t_0$, $\rho_0 v_0^2$, $v_0^2$ and $\sqrt{4\pi\rho_0}v_0$ respectively. The ideal MHD equations are solved using the quasi-monotone TVD scheme, which has the third order of approximation in the spatial variable in regions of a smooth solution and the first order of approximation in time. The scheme guarantees the conservation of mass, momentum and energy on the computational grid. Poisson's equation for the gravitational potential is solved with the implicit alternating direction method. The divergence cleaning of the magnetic field is implemented using the generalized method of Lagrange multipliers. The total number of grid cells is $N_r\times N_z = 150 \times 150$.

To investigate the conditions of the formation and properties of primary disks, 19 simulation runs were performed with initial parameters in a wide range: $0.1 \le \varepsilon_{\rm t} \le 0.9$, $0 \le \varepsilon_{\rm m} \le 0.7$, which corresponds to the initial concentration in range $10^3 \le n_0 \le 10^6$~cm$^{-3}$, initial magnetic field strength $0 \le B_0 \le 10^{-4}$~G. The modeling was carried out until the formation of an opaque region in the center of the cloud. The optical thickness with respect to its own thermal radiation was calculated for the constant opacity of dust with a characteristic value $\kappa = 0.1$~cm$^2$\,g$^{-1}$~\citep{semenov03}.

\section{Fiducial run}
\label{sect:fiduc_m}

Consider the collapse of the magnetic non-rotating PSC with mass of $10\,M_{\odot}$ for the initial parameters $\varepsilon_{\rm t} = 0.3$, $\varepsilon_{\rm m} = 0.2$. The corresponding initial radius, concentration and of magnetic field strength are $R_0=0.1$~pc, $n_0=4 \times 10^4$~cm$^{-3}$, $B_0 = 4.7 \times 10^{-5} $~G, respectively. The characteristic collapse time, taking into account the influence of the electromagnetic force, can be determined as~\cite[][]{DudSaz1982}
\begin{equation}
t_{{\rm fm}} = t_{\rm ff}(1-\varepsilon_{\rm m})^{-1/2},\quad t_{\rm ff}=(3\pi/32G\rho_0)^{1/2},
\end{equation}
where $t_{\rm ff}$ is the free-fall time. For the adopted parameters, $t_{{\rm ff}} = 0.170$~Myr,  $t_{{\rm fm}} = 0.193$~Myr.

\begin{figure*}
   \centering
  \includegraphics[width=0.79\textwidth]{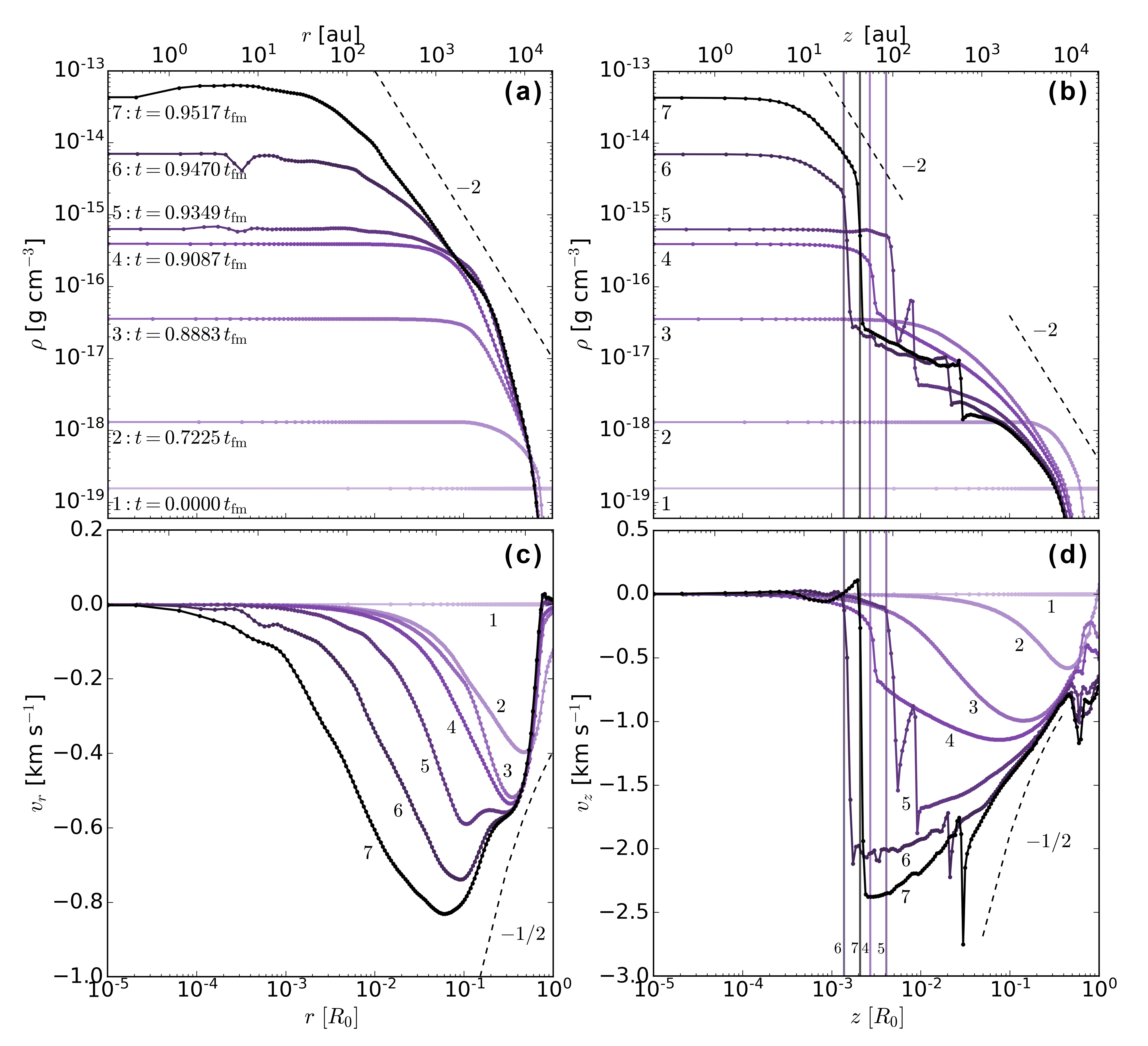}
   \caption{Left: the profiles of density (a) and radial velocity (c) in the equatorial plane at the different moments of time for the simulation with parameters $\varepsilon_{\rm t} = 0.3$, $\varepsilon_{\rm m} = 0.2$. Right: corresponding profiles of density (b) and vertical velocity (d) along the $z$-axis. Dashed lines with numbers show typical slopes, vertical lines indicate the half-thickness of the primary disk.} 
   \label{fig:profiles_t3m2w0}
\end{figure*}

In Fig.~\ref{fig:profiles_t3m2w0}, we plot the profiles of density and components of poloidal velocity vector along the $r$- and $z$-axes at several moments of time. The initial state of the considered uniform cloud is shown in this figure by the lines indicated by the number~`1'. The initial dynamics of the collapsing magnetic cloud is characterized by the propagation of a fast MHD rarefaction wave from the cloud boundary to its center~\citep{dudorov03}. The front of this wave divides the cloud into the uniform central region and nonuniform envelope. For example, line `2' in Fig.~\ref{fig:profiles_t3m2w0}a shows the rarefaction wave in the region $(r,\,z) > 0.1\,R_0$. The cloud flattens along the direction of the initial magnetic field, as shown by lines `3': the size of the central uniform region in the $z$-direction is smaller than in the $r$-direction. The outer part of the cloud is characterized by the self-similar density profile $\rho\propto r^{-2}$, in agreement with classical result of~\cite{larson1969}.

By the time $ t = 0.9087 \, t_{{\rm fm}} $ (line `4' in Fig.~\ref{fig:profiles_t3m2w0}), the density in the cloud’s center increases up to $\sim 4\times 10^{-16}$~g~cm$^{-3}$, i.e., more than three orders of magnitude in comparison with the initial value. In contrast to monotonic density decrease along the $r$-direction at the periphery of the cloud, there is an abrupt transition from the envelope to the central part of the cloud at $z = Z_{{\rm pd}} \approx 3\times 10^{-3}\,R_0\approx 60$~au along the $z$-direction. This density jump corresponds to the transition from the collapsing envelope to the quasi-magnetostatic primary disk, which will be discussed in detail below. The primary disk forms after focusing and reflection of the MHD rarefaction wave from the center of the cloud.

Later, at $t>0.9087\, t_{{\rm fm}}$ (lines `5'--`7' in Fig.~\ref{fig:profiles_t3m2w0}), another jump in the  $\rho(z)$ profiles above the primary disc is observed, $z\approx (0.01-0.03)\,R_0 \approx (200-600)$~au. This jump corresponds to the front of the fast MHD shock wave propagating outward from the disk practically in the $z$-direction. This wave forms as a result of the reflection of the fast MHD rarefaction wave from the center of the cloud, as predicted by~\cite{DudSaz1982}. The reflected fast magnetosonic wave propagates from the primary disk into the envelope, becomes steeper and turns into the shock wave. The gas is compressed by an order of magnitude in density when passing through the shock front. At the moment $t=0.9517\, t_{{\rm fm}}$, the density in the cloud center grows up to $4.3 \times 10^{-14}$~g~cm$^{-3}$, i.e., nearly five orders of magnitude in comparison with the initial value. 

Fig. 1~\ref{fig:profiles_t3m2w0}b shows that the vertical density profile above the MHD shock wave front, $z>0.05\,R_0\approx 1000$~au, also corresponds to self-similar power law $\rho \propto z^{-2}$ of the free falling gas in the envelope.

The MHD rarefaction wave propagates from the cloud boundary to its center, and the initial dynamics of the collapse is characterized by a monotonically accelerating supersonic flow in the outer part of the cloud and a decelerating flow in the inner uniform region, as Fig.~\ref{fig:profiles_t3m2w0}c shows (see lines `1'--`3'). The electromagnetic force counteracts the collapse in the $r$-direction, and the flow along the magnetic field lines along the $z$-direction is practically free. Therefore, $v_r<v_z$, as comparison of Fig~\ref{fig:profiles_t3m2w0}c and~\ref{fig:profiles_t3m2w0}d. For example, maximal radial speed is of $(-0.5)$~km~s$^{-1}$ at $r\approx 0.3\,R_0\approx 0.03$~pc (line `3'), while maximal vertical speed is of  $(-1.1)$~km~s$^{-1}$ at $z\approx 0.1\,R_0\approx 0.01$~pc. The picture changes qualitatively at   $t>0.91\, t_{{\rm fm}}$ (lines `4'--`7'). The collapse in the radial direction becomes slower than at the beginning, with a maximum speed of $v_r\approx 0.3-0.5$~km~s$^{-1}$ close to the sound speed  $v_{\rm s}=0.27$~km~s$^{-1}$. Collapse in  the $z$-is characterized by a very fast fall with the speed up to $2.5$~km~s$^{-1}$ in the envelope of the cloud, $z>Z_{\rm pd}$, and quasi-magnetostatic equilibrium, $v_z\approx 0$, in the region $z\leq Z_{\rm pd} \approx 60$~au. Equilibrium is established approximately at $t=0.9087\,t_{\rm fm}$ (line `4'). The region of rapid gas deceleration corresponds to the vertical boundary of the primary disk, which is also visible in the density profiles in Fig.~\ref{fig:profiles_t3m2w0}b. The fact that the gas falls onto the disk at a supersonic speed means that the boundary of the primary disk is a shock front. Sharp velocity peaks at $z\approx 0.01\,R_0\approx 200$~au (line `5' in Fig.~\ref{fig:profiles_t3m2w0}d), $0.02\,R_0\approx 400$~au (line `6') and $0.02\,R_0\approx 600$~au (line `7') correspond to the fast MHD shock wave propagating outside the disk. The outermost part of the cloud, $(r,\,z) > 0.1\,R_0$, characterized by the free fall velocity profile $v\propto r^{-1/2}$.

\begin{figure*}
   \centering
  \includegraphics[width=0.79\textwidth]{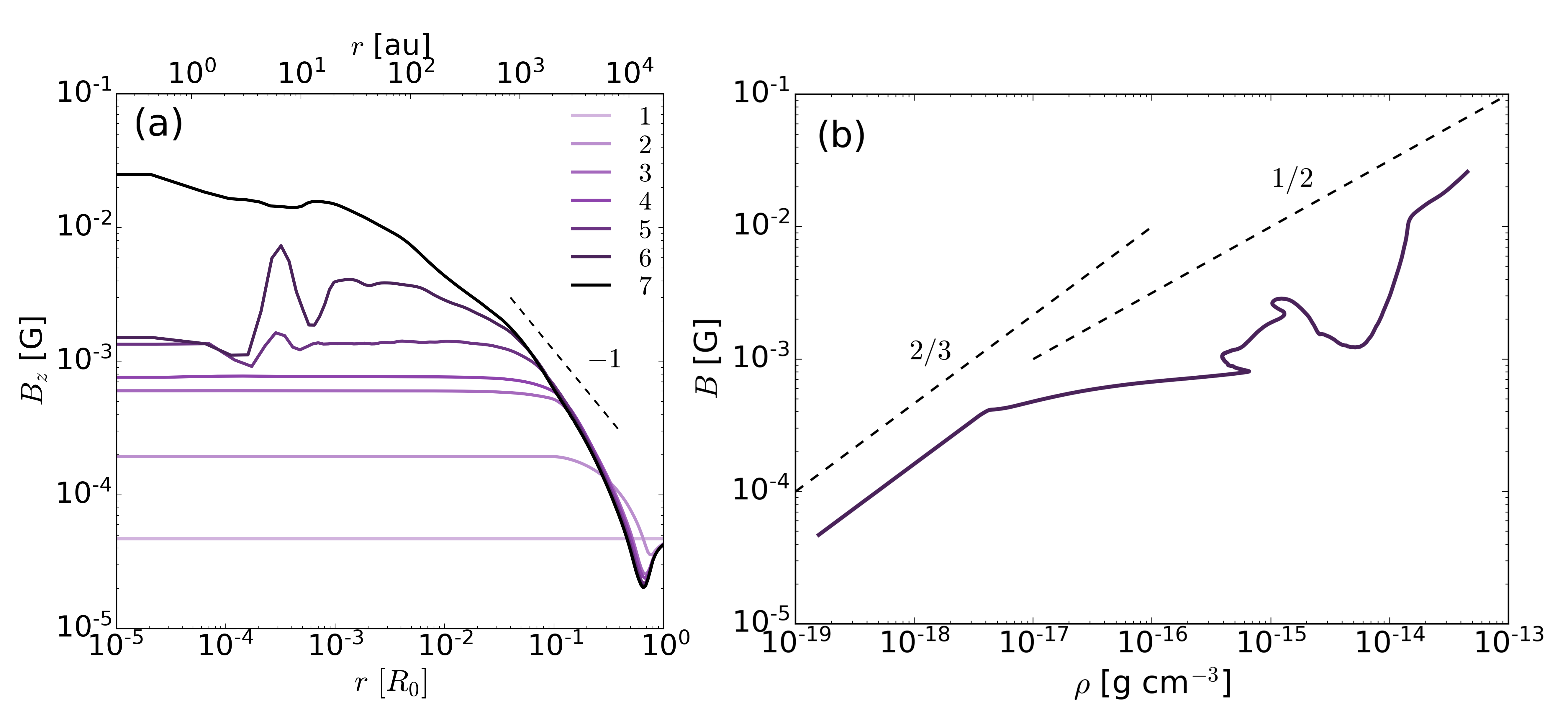}
   \caption{Left: radial profiles of  $B_z(z=0)$ at the same time moments, as in Fig.~\ref{fig:profiles_t3m2w0}. Right: the dependence of the magnetic field strength on density in the center of the cloud. Dashed lines with numbers show typical slopes. } 
   \label{fig:B_t3m2w0}
\end{figure*}

In Fig.~\ref{fig:B_t3m2w0}a we plot the radial profiles $B_z(z=0)$ at the same moments of time, as in Fig.~\ref{fig:profiles_t3m2w0}. Fig.~\ref{fig:B_t3m2w0}a shows that the magnetic field strength increases monotonically from the initial value  $4.7\times 10^{-5}$ up to $\sim 10^{-3}$~G inside the uniform part of the cloud before the formation of the primary disk (lines `1'–-`4'). The outer part of the cloud is characterized by the radial profile $B\propto r^{-1}$, which agrees with the density profile $\rho\propto r^{-2}$, in the case of the magnetostatic contraction, $B\propto \rho^{1/2}$. After the formation of the primary disk, the magnetic field strength continues to increase, but the central part of the cloud is characterized by non-monotonic profiles $B_z(r)$ (lines `5'--`7'). Local peaks in the profiles $B_z(r)$ are caused by the `bounce' of some part of the falling gas after focusing the rarefaction wave in the center of the cloud. A similar `bounce' was observed in the simulations of~\cite{scott1980} and~\cite{black1982}.

The dependence of the magnetic field strength on the gas density in the center of the cloud is shown in Fig.~\ref{fig:B_t3m2w0}b. The dependence is power law, $B\propto \rho^k_{\rm B}$, with $k_{\rm B}=2/3$ at the very early stage of the collapse, $\rho_{\rm c} < 3\times 10^{-18}\,${\rm g\,cm}$^{-3}$. This dependence reflects almost isotropic collapse of the cloud due to the weak influence of the electromagnetic force on the gas dynamics. This stage corresponds to the lines `1'--`2' in Fig.~\ref{fig:profiles_t3m2w0} and \ref{fig:B_t3m2w0}a. The dependence is almost flat, $B\approx$~const, i.e. the magnetic field strength practically does not change, in the density range $3\times 10^{-18} < \rho_{\rm c} < 5\times 10^{-16}\,${\rm g\,cm}$^{-3}$. This effect is explained by the fact that the magnetic field slows down the collapse in the radial direction, and the collapse continues mainly in the direction along the magnetic field lines, which corresponds to the case $v_r < v_z$ (lines `3'--`4' in Fig.~\ref{fig:profiles_t3m2w0}c and~\ref{fig:profiles_t3m2w0}d). Non-monotonic behavior of the $B_{\rm c}(\rho_{\rm c})$ dependence at $\rho_{\rm c}\sim (10^{-15} - 10^{-14})\,${\rm g\,cm}$^{-3}$ is caused by short-term episodes of `bounce' after focusing the rarefaction wave in the center of the cloud and its subsequent reflection. After the formation of the primary disk, $\rho_{\rm c}>10^{-14}\,${\rm g\,cm}$^{-3}$, the power law dependence $B_{\rm c}(\rho_{\rm c})$ is characterized by the index $k_{\rm B}\approx 1/2$, reflecting quasi-magnetostatic disk contraction.

\begin{figure*} 
   \centering
   \includegraphics[width=0.79\textwidth]{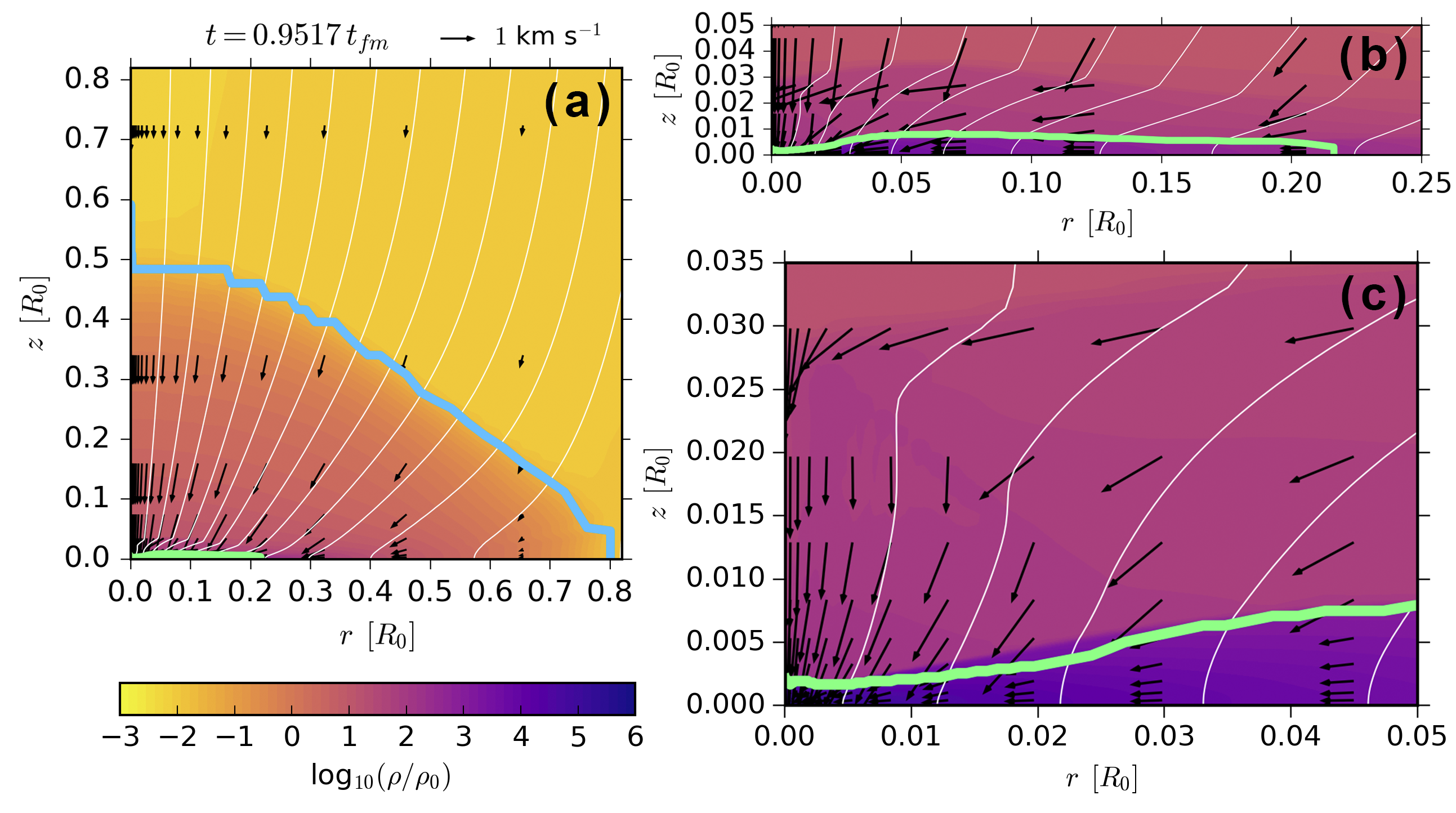}
   \caption{Two-dimensional structure of the collapsing protostellar cloud at $t = 0.9517 \, t_{\rm fm}$ for parameters $\varepsilon_{\rm t} = 0.3$, $\varepsilon_{\rm m} = 0.2$. The color filling shows the distribution of the decimal logarithm of the dimensionless density, arrows show the velocity field, white lines show the lines of the poloidal magnetic field. Panel (a): region $(r\times z) = (0.8\times 0.8)\,R_0$; panels (b) and (c): zoomed-in regions $(r\times z) = (0.25\times 0.05\,R_0)$ and $(r\times z) = (0.02\times 0.035\,R_0)$, respectively. The blue line in the panel (a) shows the cloud border, the green line in the panels (b) and (c) indicates the border of the primary disk.} 
   \label{fig:2D_t3m2w0}
\end{figure*}

In Fig.~\ref{fig:2D_t3m2w0} we plot the two-dimensional structure of the cloud at $t = 0.9517$~$t_{{\rm fm}}$, when the opaque region appears in the center of the cloud (see line `7' in Fig~\ref{fig:profiles_t3m2w0}). The distributions of the decimal logarithm of the density, vectors of the poloidal velocity and poloidal magnetic field lines in the region $(r,z)<0.8\,R_0$ (panel a) and in the central parts of the cloud (panels b and c) are depicted.

Fig.~\ref{fig:2D_t3m2w0}a demonstrates that the cloud acquires the flattened form with a semi-minor axis directed along the initial magnetic field ${\mathbf B}_0$ (see orange region in Fig~\ref{fig:2D_t3m2w0}a). The major and minor semiaxes of the flattened cloud are $R_{{\rm env}} \approx 0.62\,R_0\approx 0.062$~pc and $Z_{{\rm env}} \approx 0.4\,R_0\approx 0.040$~pc respectively. The flattening degree of the cloud $\epsilon_{{\rm env}} = Z_{{\rm env}} / R_{{\rm env}} \approx 0.645$. The magnetic field in the outer part of the envelope of the cloud, $r>(0.2-0.3)\,R_0$, has quasi-radial geometry, $B_r \sim B_z$.

In Figs~\ref{fig:2D_t3m2w0}b and \ref{fig:2D_t3m2w0}c the geometrically thin disk is seen inside the flattened cloud envelope. The distribution of the poloidal velocity arrows indicates that the rapid and almost vertical fall of the gas in the envelope turns into the slow, almost radial motion inside the disk, i.e., the quasi-magnetostatic equilibrium establishes inside the primary disk. The boundary of the primary disk is determined by the transition from vertically free-falling of matter to the magnetostatic equilibrium, characterized by $v_z\approx 0$ (see discussion of Fig.~\ref{fig:profiles_t3m2w0} above). The boundary of the primary disk determined in this way is indicated in Fig.~\ref{fig:2D_t3m2w0} with a green line. The primary disc has the radius of $R_{\rm pd}\approx 0.22\,R_0\approx 4500$~au and maximal half-thickness of $\approx 0.009\,R_0\approx 180$~au. The thickness of the primary disc is minimal along the axis of rotation, $Z_{\rm pd}\approx 0.002\,R_0\approx 40$~au. In this region, the characteristic scale height of the primary disc $H$ is of $16$~au, and 26 out of 150 computational cells fit inside the primary disk (see line `7' in Fig~\ref{fig:profiles_t3m2w0}b). The maximum cell size inside the disk is $\Delta z_{\rm max}\approx 1.6$~au~$\ll 1\,H$. Therefore, the mesh resolution is sufficient to study the internal structure of the primary disk. 

In Fig.~\ref{fig:2D_t3m2w0}b another sharp jump in density is observed at $z \approx (0.02-0.04)\,R_0\approx (400-800)$~au. This jump corresponds to the front of the fast MHD shock wave propagating from the disk into the envelope of the cloud, as shown in Fig~\ref{fig:profiles_t3m2w0}.

At $t = 0.9517$~$t_{{\rm fm}}$ the density at the center of the cloud is $4.3 \times 10^{-14}$~g~cm$^{-3}$. Further evolution of the system will lead to the formation of an almost spherical opaque hydrostatic core with $\gamma_{\rm eff}=5/3$. Accurate study of the formation of the first core and the subsequent evolution of the cloud and primary disk will be carried out in further works. 

Summarizing the results described above, we can outline the following picture of the collapse of the magnetic PSC. Initially, the collapse of the cloud is almost spherically symmetric since the electromagnetic force is weaker than the gravitational force. In the process of the collapse, the electromagnetic force increases, and eventually two-dimensional collapse turns to the one-dimensional one. The contraction of the cloud in the direction perpendicular to the magnetic field lines is slowed down by the electromagnetic force, while the vertical fall along the magnetic field lines remains practically free. One-dimensional collapse along the magnetic field lines is ineffective in the sense that this process ultimately always stops and leads to the formation of the quasi-magnetostatic equilibrium structure. Equilibrium is established due to the balance between gravity ($g_z\sim \rho GMH/r^3$, where $H$ is the scale height), gas pressure ($\partial P/\partial z \sim \rho c_{\rm T}^2/H$, where $c_{\rm T}$ is the isothermal sound speed) and electromagnetic force ($\partial(B^2/8\pi)/\partial z \sim B^2/8\pi H$) in the vertical direction, as well as balance between the gravity ($g_r\sim \rho GM/r^2$) and electromagnetic force ($\partial(B^2/8\pi)/\partial r\sim B^2/8\pi r$) in the radial direction. This force balance leads to the ratio $B\propto \rho^{1/2}$, which is observed in our simulations after the formation of the primary disk (see Fig~\ref{fig:B_t3m2w0}b). Thus, the formation of the quasi-magnetostatic primary disk at the isothermal stage is a natural and inevitable consequence of the collapse of the magnetic PSC.

\section{Evolution of the primary disk}
\label{sect:evol}

Table~\ref{tab:pd_m} lists the characteristics of the primary disk and its envelope for the parameters $\varepsilon_{\rm t} = 0.3$, $\varepsilon_{\rm m} = 0.2$ at the different time moments. The table contains the formation time (column 2), radius (column 3), maximal half-thickness (column 4), the flattening degree $\epsilon_{\rm pd} = Z_{\rm pd} / R_{\rm pd}$ (column 5) and the primary disk’s mass (column 6), the envelope’s mass (column 7), density (column 8) and magnetic field strength (column 9) at the center of the disk. The times given in rows 1--4 correspond to lines `4'–-`7' in Fig.~\ref{fig:profiles_t3m2w0}. Table~\ref{tab:pd_m} shows that the radius, $R_{\rm pd}$, and half-thickness, $Z_{\rm pd}$, of the primary disk increase with time. By the time of the opaque region formation, $R_{\rm pd}\approx 4500$~au and $Z_{\rm pd}\approx 180$~au, which corresponds to the flattening degree $\sim 0.04$. The primary disk remains geometrically and optically thin during its evolution. The growth of the primary disk with time is explained by the inflow of the matter from the envelope and corresponding increase of the quasi-magnetostatic region, characterized by a balance between gas and magnetic pressures and gravity in the $z$-direction.  
It should be noted that the half-thickness of the primary disk depends on $r$, based on our simulations. It is smaller in the central part of the disk, $r\approx 40-50$~au, as compared to its periphery, $r\approx 200$~au. This is because the magnetic field remains practically uniform in the central part of the cloud, so that the magnetic field does not make a significant contribution to the balance of forces in the vertical direction in the region $r<200$~au. Thus, the primary disk consists of the quasi-hydrostatic central part and the quasi-magnetostatic periphery. The density of the gas and the magnetic field strength at the center of the disk increase with time, reflecting the growth of the primary disk due to the inflow of the matter from the envelope.

\begin{table*}
\centering
\caption[]{Time dependence of the characteristics of the envelope of the cloud and the primary disk calculated with the initial parameters  $\varepsilon_{\rm t} = 0.3$, $\varepsilon_{\rm m} = 0.2$.\label{tab:pd_m}}
\setlength{\tabcolsep}{2.5pt}
\small
 \begin{tabular}{ccccccccc}
  \hline\noalign{\smallskip}
No & $t_{{\rm pd}}$ $[t_{\rm fm}]$ & $R_{{\rm pd}}$ $[\mbox{au}]$ & $Z_{{\rm pd}}$ $[\mbox{au}]$ & $\epsilon_{\rm pd}$ & $M_{\rm pd}\, [M_0]$ & $M_{\rm env}\, [M_0]$ & $\rho_{\rm c}$ $[\mbox{g}$~$\mbox{cm}^{-3}]$ & $B_{\rm c}$ $[\mbox{G}]$ \\
 (1) & (2) & (3) & (4) & (5) & (6) & (7) & (8) & (9) \\
  \hline\noalign{\smallskip}
1 & 0.909 & 1600 & 60 & 0.037 & 0.03 & 0.92 & $3.95\times 10^{{\rm -16}}$ & $7.59\times 10^{{\rm -4}}$ \\ 
2 & 0.935 & 3780 & 170 & 0.045 & 0.28 & 0.66 & $6.33\times 10^{{\rm -16}}$ & $7.99\times 10^{{\rm -4}}$ \\
3 & 0.947 & 4280 & 170 & 0.040 & 0.36 & 0.56 & $7.04\times 10^{{\rm -15}}$ & $1.50\times 10^{{\rm -3}}$  \\
4 & 0.952 & 4460 & 180 & 0.040 & 0.38 & 0.53 & $4.31\times 10^{{\rm -14}}$ & $2.60\times 10^{{\rm -2}}$  \\
  \noalign{\smallskip}\hline
\end{tabular}
\end{table*}

\begin{figure}
   \centering
  \includegraphics[width=0.45\textwidth]{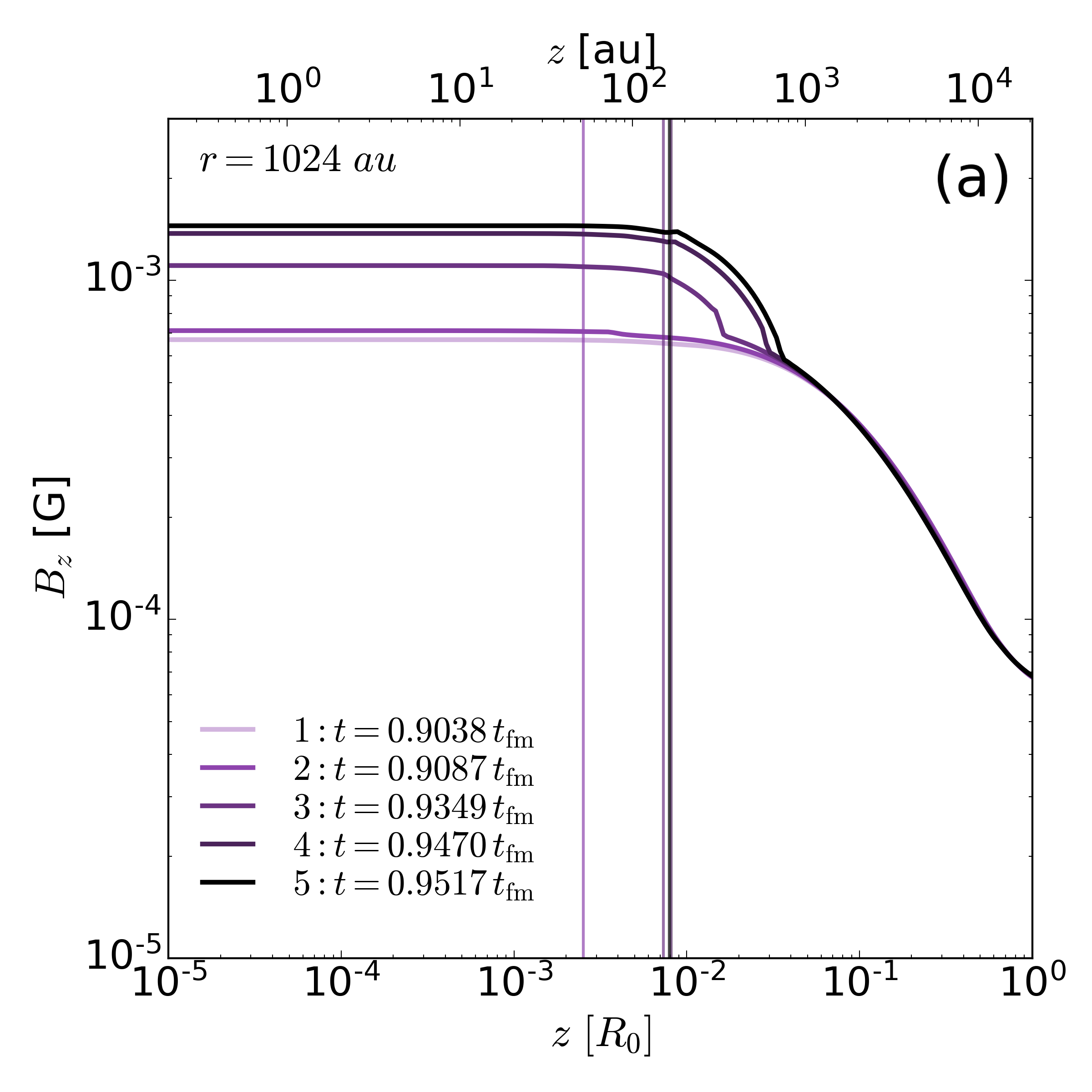}
  \includegraphics[width=0.45\textwidth]{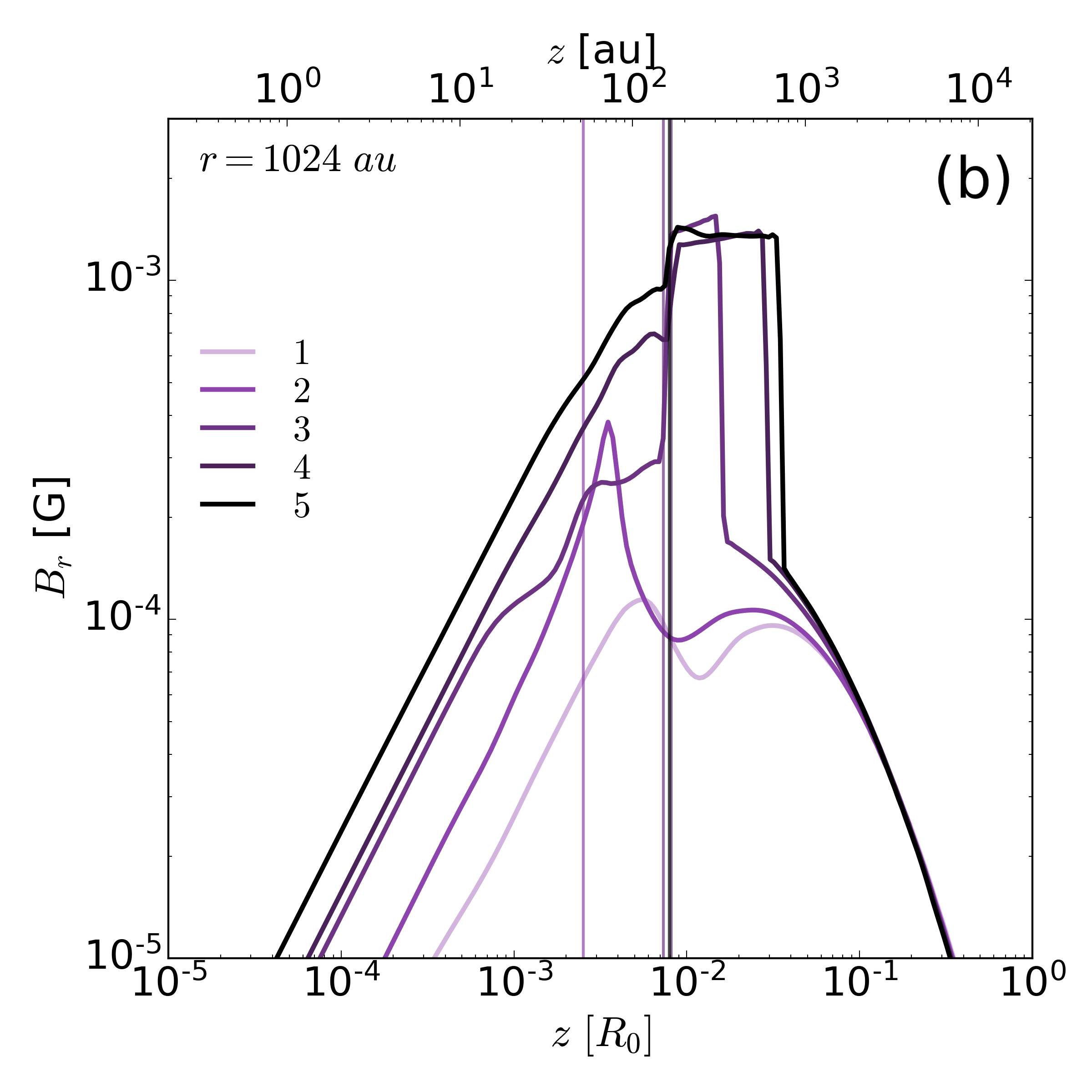}
   \caption{Panel (a): vertical profiles of $B_z$ at $r=0.05\,R_0\approx 1000$~au at several time moments, panel (b): corresponding profiles of $B_r$. Lines `2'--`5' correspond to rows 1–-4 in Table~\ref{tab:pd_m}.}
   \label{fig:BrBz_t3m2w0}
\end{figure}

In Fig.~\ref{fig:BrBz_t3m2w0} we plot the vertical profiles of $B_z$ and $B_r$ at the distance  $r=0.05\,R_0\approx 1000$~au at several time moments after the formation of the primary disk. Fig.~\ref{fig:BrBz_t3m2w0}a shows that the intensity of $B_z$ is maximal and increases with time inside the primary disk from $6\times 10^{-4}$~G at $t=0.9038\, t_{{\rm fm}}$ up to $1.5\times 10^{-3}$~G at $t=0.9517\, t_{{\rm fm}}$. The magnetic field remains uniform, $B_z\sim $~const, inside the primary disk, $z<Z_{\rm pd}$, since the gas dynamics in this region is practically one-dimensional, $v_z\ll v_r$. The area above the disk is characterized by two-dimensional collapse with $v_r\sim v_z$ (see the velocity field in Figs.~\ref{fig:2D_t3m2w0}b and~\ref{fig:2D_t3m2w0}c), the combination of radial and vertical fall causes the decrease of $B_z$ with height $z$. 

According to Fig.~\ref{fig:BrBz_t3m2w0}b, the radial magnetic field has a local maximum near the boundary of the primary disk before the fast MHD wave comes out from the disk. For example, $B_r(z=80\,\mbox{au})\approx 2.5\times 10^{-4}$~G at $t=0.9038\, t_{{\rm fm}}$ (line `1'). Component $B_r$ is maximal near the disk surface because this region is characterized by the maximal radial velocity (see Fig.~\ref{fig:2D_t3m2w0}b) and, therefore, the most efficient generation of $B_r$. The MHD shock wave is formed at the boundary of the primary disk at $t\approx 0.9087\, t_{{\rm fm}}$ (line `2'), when the reflected MHD rarefaction wave is refracted at the shock front near the disk surface and steepens as it leaves the dense disk into the rarefied envelope. Behind the wave front, there is a simultaneous increase in the gas density and the strength of the tangential component of the magnetic field (see Figs.~\ref{fig:profiles_t3m2w0}b and~\ref{fig:BrBz_t3m2w0}b), that is, this wave is a fast MHD shock wave. Thus, the magnetic field acquires a quasi-radial geometry, $B_r\sim B_z$, in the region between the disk surface and the shock front during its propagation in the disk’s envelope $z>(50-200)$~au (lines `3'--`5'). This region grows over time.

\section{Dependence on parameters}
\label{sect:param}

Let us summarize the set of simulations performed and discuss the dependence of the properties of the primary disk on the initial parameters of the cloud.

Table~\ref{tab:results_mpd10} lists the characteristics of the envelope and the primary disk in the simulations with various initial magnetic and thermal parameters. The results for two runs in which the primary disk did not form are also presented for comparison (rows 1, 2). Table~\ref{tab:results_mpd10} lists the characteristics of the structures formed at the end of the simulations. The following quantities are shown: thermal and magnetic parameters (columns 2 and 3), dimensionless density at the center of the cloud (column 4), the radii of the primary disk and envelope (columns 5 and 6, respectively), the corresponding masses (columns 7 and 8) and the flattening degrees (columns 9 and 10) of these structures, the times of the formation of the primary disk (column 11) and the opaque region ($t_{\rm od}$, column 12). Time is measured in the dynamical time scale units $t_{\rm fm}$, determined taking into account the influence of the magnetic field (see definition in the first paragraph of Section~\ref{sect:fiduc_m}).

Simulations show that the hierarchical structure `flattened envelope $\rightarrow$ primary disk $\rightarrow$ opaque region', discussed in Section~\ref{sect:fiduc_m}, forms in the simulations with $\varepsilon_{\rm m}\ge 0.2$. By the time of the opaque region formation, the density in the center of the cloud $\rho_{{\rm c}}$ increases by 5--6 orders of magnitude in comparison with the initial value. The primary discs radii vary from $0.22\,R_0$ ($\varepsilon_{\rm m}=0.2$) to $0.54\,R_0$ ($\varepsilon_{\rm m}=0.6$). 
The masses of the primary disks are 38--71~\% of the initial mass of the cloud, $M_{\rm pd} \approx 3.75-6.79\,M_{\odot}$, depending on the initial parameters of the PSCs. There is a tendency towards an increase in the size and mass of the primary disk with an increase in the magnetic parameter. 

The flattening degree of each structure in the table is defined as the ratio of the maximum half-thickness to the radius, $\epsilon = Z / R$.  The opaque region is a very thin structure with $\epsilon_{{\rm od}} \sim 10^{-4} -10^{-3} $. Subsequently, a spherically symmetric optically thick first hydrostatic core will form in its center. The primary discs are also geometrically thin, $\epsilon_{{\rm pd}} \sim 10^{-2} -10^{-1}$, and their thickness increases over time. The oblate envelope has $\epsilon_{{\rm env}}\approx 0.22-0.95$, i.e., it is geometrically thick. Therefore, the flattening degree in the central parts of the collapsing PSC is greater than on its periphery, $ \epsilon_{{\rm od}}  <  \epsilon_{{\rm pd}} < \epsilon_{{\rm env}}$.

The primary disks usually form within one dynamical timescale $t_{\rm fm}$ after the onset of the collapse in the simulations with weak or moderate magnetic field, $\varepsilon_{\rm m}<0.6$. In the cases of strong magnetic field, $\varepsilon_{\rm m}\geq 0.6$, the primary disks form earlier, at  $t\sim 0.7\,t_{\rm fm}$.

Comparison of rows ($\varepsilon_{\rm t}=0.3$, $\varepsilon_{\rm m}=0.2$) and 5 ($\varepsilon_{\rm t}=0.7$, $\varepsilon_{\rm m}=0.2$) shows that the radius, mass, and the flattening degree of the primary disk increases with increasing thermal parameter. 

\begin{table*}
\centering
\caption[]{Characteristics of the envelope of the cloud and the primary disk in the simulations of the PSC collapse with mass of $10\,M_{\odot}$.}\label{tab:results_mpd10}
\setlength{\tabcolsep}{2.5pt}
\small
 \begin{tabular}{cccccccccccc}
  \hline\noalign{\smallskip}
No &  $\varepsilon_{t}$ &  $\varepsilon_{m}$ & $\rho_{{\rm c}} / \rho_0 $ & $R_{{\rm pd}}$ $[{\rm R_0}]$ & $R_{{\rm env}}$ $[{\rm R_0}]$ & $M_{\rm pd}\, [M_0]$ & $M_{\rm env}\, [M_0]$ &  $\epsilon_{\rm pd}$ & $\epsilon_{\rm env}$ & $t_{\rm pd}$ $[t_{\rm fm}]$ & $t_{\rm od}$ $[t_{\rm fm}]$\\
 (1) & (2) & (3) & (4) & (5) & (6) & (7) & (8) & (9) & (10) & (11) & (12)\\
  \hline\noalign{\smallskip}
1 & 0.3 & 0 & $9.8\times 10^3$ & - & 0.57 & - & 0.947 & - & 1.0 & - & 0.991\\ 
2 & 0.3 & 0.01 & $1.2\times 10^4$ & - & 0.60 & - & 0.946 & - & 1.0 & - & 0.986\\
3 & 0.3 & 0.2 & $2.7\times 10^5$ & 0.22 & 0.80 & 0.38 & 0.563 & 0.04 & 0.70 & 0.909 & 0.952\\
4 & 0.3 & 0.6 & $4.6\times 10^5$ & 0.54 & 0.98 & 0.71 & 0.237 & 0.13 & 0.29 &  0.692 & 0.823\\
5 & 0.7 & 0.2 & $7.6\times 10^6$ & 0.31 & 0.83 & 0.47 & 0.470  & 0.23 & 0.63 & 1.089 & 1.194\\
  \noalign{\smallskip}\hline
\end{tabular}
\end{table*}

For comparison, in this work we also performed simulations of the collapse of magnetic PSC of solar mass with the thermal parameter  $\varepsilon_{\rm t}=0.3$ and the magnetic parameter in the range from 0.01 to 0.7. The characteristics of the structures formed at the end of the isothermal collapse of solar mass clouds are listed in Table~\ref{tab:results_mpd1}, in the same way as in Table~\ref{tab:results_mpd10}. A dash `-' means that the primary disk did not form in the corresponding run. The table confirms the conclusions drawn for a cloud with a mass $10\,M_\odot$ that the primary disk is formed under condition $\varepsilon_{\rm m} \geq 0.2$. The radius of the primary disk formed as a result of the collapse of the PSC of solar mass varies from $0.20\,R_0$ for $\varepsilon_{\rm m} = 0.2$ (row 6) to $0.68\,R_0$ for $\varepsilon_{\rm m} = 0.7$ (row 10). Comparison of these results with the results presented in Table~\ref{tab:results_mpd10}, shows that the ratio of the radius of the primary disk to the initial radius of the cloud does not depend on the cloud mass, which can be interpreted as a self-similar behavior. In the case of a dynamically strong magnetic field, $\varepsilon_{\rm m}\geq 0.5$ (rows 4 in Table~\ref{tab:results_mpd10} and 8--10 in Table~\ref{tab:results_mpd1}), the radius and mass of the primary disk are more than 50~\% of the initial radius and mass of the cloud. It can be concluded that the cloud as a whole evolves into a state of magnetostatic equilibrium when its initial magnetic field is dynamically strong, which agrees with the predictions of~\cite{DudSaz1982}.

\begin{table*}
\centering
\caption[]{Characteristics of the envelope and the primary disk in the simulations of the collapse of the PSC with mass of $1\,M_{\odot}$ for the initial thermal parameter $\varepsilon_{t}=0.3$ and various magnetic parameters $\varepsilon_{m}$.}\label{tab:results_mpd1}
\setlength{\tabcolsep}{2.5pt}
\small
 \begin{tabular}{cccccccccccc}
  \hline\noalign{\smallskip}
No & $\varepsilon_{m}$ & $\rho_{{\rm c}} / \rho_0 $ & $R_{{\rm pd}}$ $[{\rm R_0}]$ & $R_{{\rm env}}$ $[{\rm R_0}]$ & $M_{\rm pd}\, [M_0]$ & $M_{\rm env}\, [M_0]$ &  $\epsilon_{\rm pd}$ & $\epsilon_{\rm env}$ & $t_{\rm pd}$ $[t_{\rm fm}]$ & $t_{\rm od}$ $[t_{\rm fm}]$\\
 (1) & (2) & (3) & (4) & (5) & (6) & (7) & (8) & (9) & (10) & (11) \\
  \hline\noalign{\smallskip}
1 & 0 & $3.0\times 10^5$ & - & 0.64 & - & 0.95 & - & 1.0 & - & 0.995\\ 
2 & 0.001 & $2.7\times 10^5$ & - & 0.61 & - & 0.95 & - & 1.0 & - & 0.995\\ 
3 & 0.01 & $4.5\times 10^4$ & - & 0.61 & - & 0.95 & - & 1.0 & - & 0.994\\
4 & 0.05 & $9.8\times 10^3$ & - & 0.66 & - & 0.95 & - & 0.95 & - & 0.970\\ 
5 & 0.1 & $4.8\times 10^3$ & - & 0.72 & - & 0.94 & - & 0.82 & - & 0.949 \\
6 & 0.2 & $2.9\times 10^4$ & 0.20 & 0.79 & 0.33 & 0.62 & 0.04 & 0.73 & 0.92 & 0.950 \\
7 & 0.4 & $7.6\times 10^3$ & 0.42 & 0.91 & 0.63 & 0.31 & 0.06 & 0.48 & 0.82 & 0.902 \\
8 & 0.5 & $5.3\times 10^3$ & 0.59 & 0.93 & 0.77 & 0.17 & 0.07 & 0.44 & 0.76 & 0.864 \\
9 & 0.6 & $4.2\times 10^3$ & 0.61 & 0.65 & 0.73 & 0.22 & 0.11 & 0.57 & 0.69 & 0.814 \\
10 & 0.7 & $4.2\times 10^3$ & 0.68 & 0.92 & 0.71 & 0.23 & 0.08 & 0.17 & 0.61 & 0.742 \\
  \noalign{\smallskip}\hline
\end{tabular}
\end{table*}

Comparison of column 7 in Table~\ref{tab:results_mpd10} and 6 in Table~\ref{tab:results_mpd1} shows that the ratio of the mass of the primary disk to the mass of the cloud is also independent of the absolute value of the initial mass of the cloud. This ratio increases from 33~\% at $\varepsilon_{\rm m} = 0.2$ up to 71~\% at $\varepsilon_{\rm m} = 0.7$. The primary disks in the case of the PSC of solar mass are also geometrically thin, but their flattening degree is slightly greater than in the case of $M_0=10\,M_{\odot}$.

\section{Conclusions and discussion}
\label{sect:end}

In this work, we performed two-dimensional numerical MHD simulations of the collapse of the magnetic PSCs with mass $M_0=10$ and $1\, M_{\odot}$ for a wide range of initial thermal and magnetic parameters, $\varepsilon_{\rm t}$ and $\varepsilon_{\rm m}$. The simulations show that the PSCs acquire the hierarchical structure at the end of the isothermal stage of the collapse. Initially spherical cloud takes the form of an oblate envelope due to the action of the electromagnetic force. The envelope is optically thin and geometrically thick with the flattening degree $Z/R\sim (0.20-0.95)$. The optically and geometrically thin quasi-magnetostatic disk with $Z/R\sim 10^{-2}-10^{-1}$ forms inside this envelope. We call this structure as the primary disk because it is the seed for the further formation of the protostellar disk. Primary discs form in the case $\varepsilon_{\rm m}\geq 0.2$. Subsequently, the first hydrostatic core will form in the center of the cloud. 

The mass and radius of the primary discs increase with increasing $\varepsilon_{\rm m}$ and lie in ranges $M_{\rm pd}\approx (0.3-0.7)\,M_0$ and $R_{\rm pd}\approx 0.2-0.7\,R_0$, respectively. Collapse of the PSCs with strong magnetic field ($\varepsilon_{\rm m} \ge 0.5$) leads to the formation of massive primary disks, the mass of which is greater than the mass of the envelope, i.e., the cloud as a whole evolves into the state of quasi-magnetostatic equilibrium, as predicted by~\cite{DudSaz1982}.

The primary disks form after focusing and reflection of the rarefaction wave in the center of the cloud, in accordance with the conclusion of~\cite{dudorov03}. This happens in $t\approx 1\,t_{{\rm fm}}$ after the beginning of the collapse of the protostellar clouds with moderate magnetic field, $\varepsilon_{\rm m}\approx 0.2$. Here $t_{{\rm fm}}$ is the characteristic dynamical time scale for the collapse, taking into account the influence of the magnetic field. In the case of the PSC with dynamically strong magnetic field, $\varepsilon_{\rm m} \geq 0.5$, the primary discs form faster, within $\sim 0.7-0.8\,t_{{\rm fm}}$.  The primary discs are long-lived structures. They grow with time until the end of the isothermal stage of the collapse in our simulations. Further evolution of the disks will be determined by the attenuation of their magnetic flux due to magnetic ambipolar diffusion and Ohmic dissipation and the subsequent formation of a protostar with a protostellar Keplerian disk. 

It can be assumed that the primary disks will appear in observations as the flattened envelopes around protostars with protostellar disks that form after the isothermal collapse. Large-scale flat structures observed in class 0 YSOs can be interpreted as the primary disks. For example, such a flattened structure was found in the emission of molecular line NH$_3$(1,1) from the HH211 YSO~\citep{wiseman2001}. This primary disk has radius of $6000$~au. Another example is the flattened envelope with radius of $2000$~au in the L1527 YSO~\citep{ohashi97}.

In recent years, more compact disks have been found around protostars in class 0 YSOs. These disks are observed in the IR continuum, as well as in the emission of CO lines using the ALMA and VLT~\citep[see, e.g.,][]{jorgensen09, enoch11, yen2017, maret2020, tobin2020}. There are indications that some of these compact discs have rotation profiles close to the Keplerian one. Compact rotating discs have also been found inside the flattened envelopes HH211 and L1527 discussed above~\cite{lee2019} and \cite{tobin2012}, respectively. Based on the performed modeling and analysis of the observations, we conclude that the primary disks determine the initial conditions for the formation of protostars with protostellar disks. Therefore, the question arises whether it is possible to observe the hierarchy including primary disks, especially in the collapsing PSCs before the formation of the first core. We assume that the envelope and the primary disk can be distinguished from each other in terms of angular momentum distribution and magnetic field geometry. The angular momentum distribution can be investigated using observations of molecular emission lines~\cite[see, e.g.,][]{punanova2018, guedel2020}. The geometry of the magnetic field can be determined using polarization mapping, as was done for HH211~\citep{lee2019}. The simulations performed in our work show that the primary disk has a quasi-uniform magnetic field inside, and a quasi-radial magnetic field near the surface of the primary disk immediately after its formation. In the process of further evolution, the fast MHD shock wave propagates beyond the primary disk, and the region of the quasi-radial magnetic field is bounded by the surface of the primary disk and the shock front.  

Further development of the work involves the investigation of the collapse of rotating magnetic PSCs, the evolution of the angular momentum of primary and protostellar disks, as well as the influence of magnetic ambipolar diffusion and Ohmic dissipation on the evolution of their magnetic flux. It is necessary to determine the total angular momentum, mass and magnetic flux of the envelope, primary disk, and opaque core in order to investigate the observational manifestations of the internal hierarchy of the collapsing cloud. An important task is to consider a more realistic initial configuration, in which the PSC with magnetic field is initially non-uniform.

It should be emphasized that the primary disks defined in this work are not pseudodisks described in the simulations of~\cite{Galli1993}, where the problem statement of the accretion stage of the collapse was considered. This problem statement corresponds to advanced stages of the collapse, when a highly nonuniform cloud structure with a sharp increase in density in the center has already formed. The simulations of this work show that flattened quasi-static disks form at the earliest stages of collapse before the formation of the first hydrostatic core. It should also be noted that the performed simulations show that the magnetic field remains quasi-uniform inside the central flattened region of the collapsing cloud, in contrast to the unphysical statement of~\cite{galli2006} that the magnetic field of a monopole type is formed in such a region.

{\bf Acknowledgments.}
The work of S. A. Khaibrakhmanov in Section \ref{sect:end} was supported by the Russian Foundation for Basic Research (project 18-52-52006). The work of A. E. Dudorov in Section  \ref{sect:fiduc_m} was supported by the Russian Foundation for Basic Research (project 18-02-01067). The work of N.~S.~Kargaltseva in Section \ref{sect:param} was supported by the Russian Science Foundation (project 19-72-10012). A. G. Zhilkin's work in Section~\ref{sect:problem} was supported by the Government of the Russian Federation and the Ministry of Higher Education and Science of the Russian Federation, grant 075-15-2020-780 (N13.1902.21.0039).

\bibliographystyle{mnras}
\bibliography{bib}

\begin{thebibliography}{}
\makeatletter
\relax
\def\mn@urlcharsother{\let\do\@makeother \do\$\do\&\do\#\do\^\do\_\do\%\do\~}
\def\mn@doi{\begingroup\mn@urlcharsother \@ifnextchar [ {\mn@doi@}
  {\mn@doi@[]}}
\def\mn@doi@[#1]#2{\def\@tempa{#1}\ifx\@tempa\@empty \href
  {http://dx.doi.org/#2} {doi:#2}\else \href {http://dx.doi.org/#2} {#1}\fi
  \endgroup}
\def\mn@eprint#1#2{\mn@eprint@#1:#2::\@nil}
\def\mn@eprint@arXiv#1{\href {http://arxiv.org/abs/#1} {{\tt arXiv:#1}}}
\def\mn@eprint@dblp#1{\href {http://dblp.uni-trier.de/rec/bibtex/#1.xml}
  {dblp:#1}}
\def\mn@eprint@#1:#2:#3:#4\@nil{\def\@tempa {#1}\def\@tempb {#2}\def\@tempc
  {#3}\ifx \@tempc \@empty \let \@tempc \@tempb \let \@tempb \@tempa \fi \ifx
  \@tempb \@empty \def\@tempb {arXiv}\fi \@ifundefined
  {mn@eprint@\@tempb}{\@tempb:\@tempc}{\expandafter \expandafter \csname
  mn@eprint@\@tempb\endcsname \expandafter{\@tempc}}}

\bibitem[\protect\citeauthoryear{{Andre}, {Ward-Thompson}  \&
  {Barsony}}{{Andre} et~al.}{1993}]{andre1993}
{Andre} P.,  {Ward-Thompson} D.,   {Barsony} M.,  1993, \mn@doi [\apj]
  {10.1086/172425}, \href
  {https://ui.adsabs.harvard.edu/abs/1993ApJ...406..122A} {406, 122}

\bibitem[\protect\citeauthoryear{{Black} \& {Scott}}{{Black} \&
  {Scott}}{1982}]{black1982}
{Black} D.~C.,  {Scott} E.~H.,  1982, \mn@doi [\apj] {10.1086/160541}, \href
  {https://ui.adsabs.harvard.edu/abs/1982ApJ...263..696B} {263, 696}

\bibitem[\protect\citeauthoryear{{Caselli}, {Benson}, {Myers}  \&
  {Tafalla}}{{Caselli} et~al.}{2002}]{caselli02}
{Caselli} P.,  {Benson} P.~J.,  {Myers} P.~C.,   {Tafalla} M.,  2002, \mn@doi
  [\apj] {10.1086/340195}, \href
  {https://ui.adsabs.harvard.edu/abs/2002ApJ...572..238C} {572, 238}

\bibitem[\protect\citeauthoryear{{Chapman} et~al.,}{{Chapman}
  et~al.}{2013}]{Chapman_2013}
{Chapman} N.~L.,  et~al., 2013, \mn@doi [\apj] {10.1088/0004-637X/770/2/151},
  \href {https://ui.adsabs.harvard.edu/abs/2013ApJ...770..151C} {770, 151}

\bibitem[\protect\citeauthoryear{{Crutcher}}{{Crutcher}}{2012}]{crutcher2012}
{Crutcher} R.~M.,  2012, \mn@doi [\araa] {10.1146/annurev-astro-081811-125514},
  \href {https://ui.adsabs.harvard.edu/abs/2012ARA&A..50...29C} {50, 29}

\bibitem[\protect\citeauthoryear{{Davidson} et~al.,}{{Davidson}
  et~al.}{2011}]{Davidson_2011}
{Davidson} J.~A.,  et~al., 2011, \mn@doi [\apj] {10.1088/0004-637X/732/2/97},
  \href {https://ui.adsabs.harvard.edu/abs/2011ApJ...732...97D} {732, 97}

\bibitem[\protect\citeauthoryear{{Dudorov} \& {Sazonov}}{{Dudorov} \&
  {Sazonov}}{1982}]{DudSaz1982}
{Dudorov} A.~E.,  {Sazonov} I.~V.,  1982, Nauchnye Informatsii, \href
  {https://ui.adsabs.harvard.edu/abs/1982NInfo..50...98D} {50, 98}

\bibitem[\protect\citeauthoryear{{Dudorov} \& {Zhilkin}}{{Dudorov} \&
  {Zhilkin}}{2003}]{dudorov03}
{Dudorov} A.~E.,  {Zhilkin} A.~G.,  2003, \mn@doi [Soviet Journal of
  Experimental and Theoretical Physics] {10.1134/1.1560390}, \href
  {https://ui.adsabs.harvard.edu/abs/2003JETP...96..165D} {96, 165}

\bibitem[\protect\citeauthoryear{{Dudorov}, {Zhilkin}  \&
  {Kuznetsov}}{{Dudorov} et~al.}{1999a}]{dudorov1999mma}
{Dudorov} A.~E.,  {Zhilkin} A.~G.,   {Kuznetsov} O.~A.,  1999a, Matem.
  Modelir., 11, 101

\bibitem[\protect\citeauthoryear{{Dudorov}, {Zhilkin}  \&
  {Kuznetsov}}{{Dudorov} et~al.}{1999b}]{dudorov1999mmb}
{Dudorov} A.~E.,  {Zhilkin} A.~G.,   {Kuznetsov} O.~A.,  1999b, Matem.
  Modelir., 11, 110

\bibitem[\protect\citeauthoryear{{Dudorov}, {Zhilkin}  \&
  {Kuznetsov}}{{Dudorov} et~al.}{1999c}]{Dud1999}
{Dudorov} A.~E.,  {Zhilkin} A.~G.,   {Kuznetsov} O.~A.,  1999c, in {Miyama}
  S.~M.,  {Tomisaka} K.,   {Hanawa} T.,  eds,  Astrophysics and Space Science
  Library Vol. 240, Numerical Astrophysics. p.~389,
  \mn@doi{10.1007/978-94-011-4780-4_116}

\bibitem[\protect\citeauthoryear{{Dudorov}, {Zhilkin}, {Lazareva}  \&
  {Kuznetsov}}{{Dudorov} et~al.}{2000}]{dudorov2000}
{Dudorov} A.~E.,  {Zhilkin} A.~G.,  {Lazareva} N.~Y.,   {Kuznetsov} O.~A.,
  2000, \mn@doi [Astronomical and Astrophysical Transactions]
  {10.1080/10556790008238597}, \href
  {https://ui.adsabs.harvard.edu/abs/2000A&AT...19..515D} {19, 515}

\bibitem[\protect\citeauthoryear{{Enoch} et~al.,}{{Enoch}
  et~al.}{2011}]{enoch11}
{Enoch} M.~L.,  et~al., 2011, \mn@doi [\apjs] {10.1088/0067-0049/195/2/21},
  \href {https://ui.adsabs.harvard.edu/abs/2011ApJS..195...21E} {195, 21}

\bibitem[\protect\citeauthoryear{{Galametz} et~al.,}{{Galametz}
  et~al.}{2018}]{galametz2018}
{Galametz} M.,  et~al., 2018, \mn@doi [\aap] {10.1051/0004-6361/201833004},
  \href {https://ui.adsabs.harvard.edu/abs/2018A&A...616A.139G} {616, A139}

\bibitem[\protect\citeauthoryear{{Galli} \& {Shu}}{{Galli} \&
  {Shu}}{1993}]{Galli1993}
{Galli} D.,  {Shu} F.~H.,  1993, \mn@doi [\apj] {10.1086/173305}, \href
  {https://ui.adsabs.harvard.edu/abs/1993ApJ...417..220G} {417, 220}

\bibitem[\protect\citeauthoryear{{Galli}, {Lizano}, {Shu}  \& {Allen}}{{Galli}
  et~al.}{2006}]{galli2006}
{Galli} D.,  {Lizano} S.,  {Shu} F.~H.,   {Allen} A.,  2006, \mn@doi [\apj]
  {10.1086/505257}, \href
  {https://ui.adsabs.harvard.edu/abs/2006ApJ...647..374G} {647, 374}

\bibitem[\protect\citeauthoryear{{Gaudel} et~al.,}{{Gaudel}
  et~al.}{2020}]{guedel2020}
{Gaudel} M.,  et~al., 2020, \mn@doi [\aap] {10.1051/0004-6361/201936364}, \href
  {https://ui.adsabs.harvard.edu/abs/2020A&A...637A..92G} {637, A92}

\bibitem[\protect\citeauthoryear{{Girart}, {Rao}  \& {Marrone}}{{Girart}
  et~al.}{2006}]{girart2006}
{Girart} J.~M.,  {Rao} R.,   {Marrone} D.~P.,  2006, \mn@doi [Science]
  {10.1126/science.1129093}, \href
  {https://ui.adsabs.harvard.edu/abs/2006Sci...313..812G} {313, 812}

\bibitem[\protect\citeauthoryear{{Hennebelle} \& {Ciardi}}{{Hennebelle} \&
  {Ciardi}}{2009}]{hennebelle2009}
{Hennebelle} P.,  {Ciardi} A.,  2009, \mn@doi [\aap]
  {10.1051/0004-6361/200913008}, \href
  {https://ui.adsabs.harvard.edu/abs/2009A&A...506L..29H} {506, L29}

\bibitem[\protect\citeauthoryear{{Hull} et~al.,}{{Hull}
  et~al.}{2014}]{Hull_2014}
{Hull} C. L.~H.,  et~al., 2014, \mn@doi [\apjs] {10.1088/0067-0049/213/1/13},
  \href {https://ui.adsabs.harvard.edu/abs/2014ApJS..213...13H} {213, 13}

\bibitem[\protect\citeauthoryear{{J{\o}rgensen}, {van Dishoeck}, {Visser},
  {Bourke}, {Wilner}, {Lommen}, {Hogerheijde}  \& {Myers}}{{J{\o}rgensen}
  et~al.}{2009}]{jorgensen09}
{J{\o}rgensen} J.~K.,  {van Dishoeck} E.~F.,  {Visser} R.,  {Bourke} T.~L.,
  {Wilner} D.~J.,  {Lommen} D.,  {Hogerheijde} M.~R.,   {Myers} P.~C.,  2009,
  \mn@doi [\aap] {10.1051/0004-6361/200912325}, \href
  {https://ui.adsabs.harvard.edu/abs/2009A&A...507..861J} {507, 861}

\bibitem[\protect\citeauthoryear{{Lam}, {Li}, {Chen}, {Tomida}  \&
  {Zhao}}{{Lam} et~al.}{2019}]{lam2019}
{Lam} K.~H.,  {Li} Z.-Y.,  {Chen} C.-Y.,  {Tomida} K.,   {Zhao} B.,  2019,
  \mn@doi [\mnras] {10.1093/mnras/stz2436}, \href
  {https://ui.adsabs.harvard.edu/abs/2019MNRAS.489.5326L} {489, 5326}

\bibitem[\protect\citeauthoryear{{Larson}}{{Larson}}{1969}]{larson1969}
{Larson} R.~B.,  1969, \mn@doi [\mnras] {10.1093/mnras/145.3.271}, \href
  {https://ui.adsabs.harvard.edu/abs/1969MNRAS.145..271L} {145, 271}

\bibitem[\protect\citeauthoryear{{Launhardt} et~al.,}{{Launhardt}
  et~al.}{2013}]{launhardt2013}
{Launhardt} R.,  et~al., 2013, \mn@doi [\aap] {10.1051/0004-6361/201220477},
  \href {https://ui.adsabs.harvard.edu/abs/2013A&A...551A..98L} {551, A98}

\bibitem[\protect\citeauthoryear{{Lee}, {Hirano}, {Zhang}, {Shang}, {Ho}  \&
  {Krasnopolsky}}{{Lee} et~al.}{2014}]{lee2014}
{Lee} C.-F.,  {Hirano} N.,  {Zhang} Q.,  {Shang} H.,  {Ho} P. T.~P.,
  {Krasnopolsky} R.,  2014, \mn@doi [\apj] {10.1088/0004-637X/786/2/114}, \href
  {https://ui.adsabs.harvard.edu/abs/2014ApJ...786..114L} {786, 114}

\bibitem[\protect\citeauthoryear{{Lee} et~al.,}{{Lee} et~al.}{2019}]{lee2019}
{Lee} C.-F.,  et~al., 2019, \mn@doi [\apj] {10.3847/1538-4357/ab2458}, \href
  {https://ui.adsabs.harvard.edu/abs/2019ApJ...879..101L} {879, 101}

\bibitem[\protect\citeauthoryear{{Li}, {Dowell}, {Goodman}, {Hildebrand}  \&
  {Novak}}{{Li} et~al.}{2009}]{li2009}
{Li} H.-b.,  {Dowell} C.~D.,  {Goodman} A.,  {Hildebrand} R.,   {Novak} G.,
  2009, \mn@doi [\apj] {10.1088/0004-637X/704/2/891}, \href
  {https://ui.adsabs.harvard.edu/abs/2009ApJ...704..891L} {704, 891}

\bibitem[\protect\citeauthoryear{{Lindberg} et~al.,}{{Lindberg}
  et~al.}{2014}]{lindberg2014}
{Lindberg} J.~E.,  et~al., 2014, \mn@doi [\aap] {10.1051/0004-6361/201322651},
  \href {https://ui.adsabs.harvard.edu/abs/2014A&A...566A..74L} {566, A74}

\bibitem[\protect\citeauthoryear{{Looney}, {Tobin}  \& {Kwon}}{{Looney}
  et~al.}{2007}]{looney2007}
{Looney} L.~W.,  {Tobin} J.~J.,   {Kwon} W.,  2007, \mn@doi [\apjl]
  {10.1086/524361}, \href
  {https://ui.adsabs.harvard.edu/abs/2007ApJ...670L.131L} {670, L131}

\bibitem[\protect\citeauthoryear{{Maret} et~al.,}{{Maret}
  et~al.}{2020}]{maret2020}
{Maret} S.,  et~al., 2020, \mn@doi [\aap] {10.1051/0004-6361/201936798}, \href
  {https://ui.adsabs.harvard.edu/abs/2020A&A...635A..15M} {635, A15}

\bibitem[\protect\citeauthoryear{{Murillo} \& {Lai}}{{Murillo} \&
  {Lai}}{2013}]{murillo2013}
{Murillo} N.~M.,  {Lai} S.-P.,  2013, \mn@doi [\apjl]
  {10.1088/2041-8205/764/1/L15}, \href
  {https://ui.adsabs.harvard.edu/abs/2013ApJ...764L..15M} {764, L15}

\bibitem[\protect\citeauthoryear{{Nakano}}{{Nakano}}{1979}]{nakano1979}
{Nakano} T.,  1979, \pasj, \href
  {https://ui.adsabs.harvard.edu/abs/1979PASJ...31..697N} {31, 697}

\bibitem[\protect\citeauthoryear{{Nakano}, {Nishi}  \& {Umebayashi}}{{Nakano}
  et~al.}{2002}]{nakano2002}
{Nakano} T.,  {Nishi} R.,   {Umebayashi} T.,  2002, \mn@doi [\apj]
  {10.1086/340587}, \href
  {https://ui.adsabs.harvard.edu/abs/2002ApJ...573..199N} {573, 199}

\bibitem[\protect\citeauthoryear{{Ohashi}, {Hayashi}, {Ho}  \&
  {Momose}}{{Ohashi} et~al.}{1997}]{ohashi97}
{Ohashi} N.,  {Hayashi} M.,  {Ho} P. T.~P.,   {Momose} M.,  1997, \mn@doi
  [\apj] {10.1086/303533}, \href
  {https://ui.adsabs.harvard.edu/abs/1997ApJ...475..211O} {475, 211}

\bibitem[\protect\citeauthoryear{{Punanova}, {Caselli}, {Pineda}, {Pon},
  {Tafalla}, {Hacar}  \& {Bizzocchi}}{{Punanova} et~al.}{2018}]{punanova2018}
{Punanova} A.,  {Caselli} P.,  {Pineda} J.~E.,  {Pon} A.,  {Tafalla} M.,
  {Hacar} A.,   {Bizzocchi} L.,  2018, \mn@doi [\aap]
  {10.1051/0004-6361/201731159}, \href
  {https://ui.adsabs.harvard.edu/abs/2018A&A...617A..27P} {617, A27}

\bibitem[\protect\citeauthoryear{{Sadavoy} et~al.,}{{Sadavoy}
  et~al.}{2018}]{sadavoy2018}
{Sadavoy} S.~I.,  et~al., 2018, \mn@doi [\apj] {10.3847/1538-4357/aaa080},
  \href {https://ui.adsabs.harvard.edu/abs/2018ApJ...852..102S} {852, 102}

\bibitem[\protect\citeauthoryear{{Scott} \& {Black}}{{Scott} \&
  {Black}}{1980}]{scott1980}
{Scott} E.~H.,  {Black} D.~C.,  1980, \mn@doi [\apj] {10.1086/158098}, \href
  {https://ui.adsabs.harvard.edu/abs/1980ApJ...239..166S} {239, 166}

\bibitem[\protect\citeauthoryear{{Semenov}, {Henning}, {Helling}, {Ilgner}  \&
  {Sedlmayr}}{{Semenov} et~al.}{2003}]{semenov03}
{Semenov} D.,  {Henning} T.,  {Helling} C.,  {Ilgner} M.,   {Sedlmayr} E.,
  2003, \mn@doi [\aap] {10.1051/0004-6361:20031279}, \href
  {https://ui.adsabs.harvard.edu/abs/2003A&A...410..611S} {410, 611}

\bibitem[\protect\citeauthoryear{{Tang} et~al.,}{{Tang}
  et~al.}{2018}]{tang2018}
{Tang} M.,  et~al., 2018, \mn@doi [\apj] {10.3847/1538-4357/aaadad}, \href
  {https://ui.adsabs.harvard.edu/abs/2018ApJ...856..141T} {856, 141}

\bibitem[\protect\citeauthoryear{{Teixeira}, {Lada}  \& {Alves}}{{Teixeira}
  et~al.}{2005}]{teixeira2005}
{Teixeira} P.~S.,  {Lada} C.~J.,   {Alves} J.~F.,  2005, \mn@doi [\apj]
  {10.1086/430849}, \href
  {https://ui.adsabs.harvard.edu/abs/2005ApJ...629..276T} {629, 276}

\bibitem[\protect\citeauthoryear{{Tobin}, {Hartmann}, {Looney}  \&
  {Chiang}}{{Tobin} et~al.}{2010}]{tobin2010}
{Tobin} J.~J.,  {Hartmann} L.,  {Looney} L.~W.,   {Chiang} H.-F.,  2010,
  \mn@doi [\apj] {10.1088/0004-637X/712/2/1010}, \href
  {https://ui.adsabs.harvard.edu/abs/2010ApJ...712.1010T} {712, 1010}

\bibitem[\protect\citeauthoryear{{Tobin}, {Hartmann}, {Chiang}, {Wilner},
  {Looney}, {Loinard}, {Calvet}  \& {D'Alessio}}{{Tobin}
  et~al.}{2012}]{tobin2012}
{Tobin} J.~J.,  {Hartmann} L.,  {Chiang} H.-F.,  {Wilner} D.~J.,  {Looney}
  L.~W.,  {Loinard} L.,  {Calvet} N.,   {D'Alessio} P.,  2012, \mn@doi [\nat]
  {10.1038/nature11610}, \href
  {https://ui.adsabs.harvard.edu/abs/2012Natur.492...83T} {492, 83}

\bibitem[\protect\citeauthoryear{{Tobin} et~al.,}{{Tobin}
  et~al.}{2020}]{tobin2020}
{Tobin} J.~J.,  et~al., 2020, \mn@doi [\apj] {10.3847/1538-4357/ab6f64}, \href
  {https://ui.adsabs.harvard.edu/abs/2020ApJ...890..130T} {890, 130}

\bibitem[\protect\citeauthoryear{{Tomisaka}}{{Tomisaka}}{2002}]{tomisaka2002}
{Tomisaka} K.,  2002, \mn@doi [\apj] {10.1086/341133}, \href
  {https://ui.adsabs.harvard.edu/abs/2002ApJ...575..306T} {575, 306}

\bibitem[\protect\citeauthoryear{{Troland} \& {Crutcher}}{{Troland} \&
  {Crutcher}}{2008}]{troland2008}
{Troland} T.~H.,  {Crutcher} R.~M.,  2008, \mn@doi [\apj] {10.1086/587546},
  \href {https://ui.adsabs.harvard.edu/abs/2008ApJ...680..457T} {680, 457}

\bibitem[\protect\citeauthoryear{{Tsukamoto}, {Okuzumi}, {Iwasaki}, {Machida}
  \& {Inutsuka}}{{Tsukamoto} et~al.}{2017}]{tsukamoto2017}
{Tsukamoto} Y.,  {Okuzumi} S.,  {Iwasaki} K.,  {Machida} M.~N.,   {Inutsuka}
  S.-i.,  2017, \mn@doi [\pasj] {10.1093/pasj/psx113}, \href
  {https://ui.adsabs.harvard.edu/abs/2017PASJ...69...95T} {69, 95}

\bibitem[\protect\citeauthoryear{{Wiseman}, {Wootten}, {Zinnecker}  \&
  {McCaughrean}}{{Wiseman} et~al.}{2001}]{wiseman2001}
{Wiseman} J.,  {Wootten} A.,  {Zinnecker} H.,   {McCaughrean} M.,  2001,
  \mn@doi [\apjl] {10.1086/319474}, \href
  {https://ui.adsabs.harvard.edu/abs/2001ApJ...550L..87W} {550, L87}

\bibitem[\protect\citeauthoryear{{Yen}, {Koch}, {Takakuwa}, {Krasnopolsky},
  {Ohashi}  \& {Aso}}{{Yen} et~al.}{2017}]{yen2017}
{Yen} H.-W.,  {Koch} P.~M.,  {Takakuwa} S.,  {Krasnopolsky} R.,  {Ohashi} N.,
  {Aso} Y.,  2017, \mn@doi [\apj] {10.3847/1538-4357/834/2/178}, \href
  {https://ui.adsabs.harvard.edu/abs/2017ApJ...834..178Y} {834, 178}

\bibitem[\protect\citeauthoryear{{Zhao}, {Caselli}, {Li}  \&
  {Krasnopolsky}}{{Zhao} et~al.}{2018}]{Zhao18}
{Zhao} B.,  {Caselli} P.,  {Li} Z.-Y.,   {Krasnopolsky} R.,  2018, \mn@doi
  [\mnras] {10.1093/mnras/stx2617}, \href
  {https://ui.adsabs.harvard.edu/abs/2018MNRAS.473.4868Z} {473, 4868}

\bibitem[\protect\citeauthoryear{{Zhao} et~al.,}{{Zhao}
  et~al.}{2020}]{zhao2020}
{Zhao} B.,  et~al., 2020, \mn@doi [\ssr] {10.1007/s11214-020-00664-z}, \href
  {https://ui.adsabs.harvard.edu/abs/2020SSRv..216...43Z} {216, 43}

\bibitem[\protect\citeauthoryear{{Zhilkin}, {Pavlyuchenkov}  \&
  {Zamozdra}}{{Zhilkin} et~al.}{2009}]{zhilkin09}
{Zhilkin} A.~G.,  {Pavlyuchenkov} Y.~N.,   {Zamozdra} S.~N.,  2009, \mn@doi
  [Astronomy Reports] {10.1134/S1063772909070026}, \href
  {https://ui.adsabs.harvard.edu/abs/2009ARep...53..590Z} {53, 590}

\makeatother
\end{thebibliography}

\end{document}